\documentclass[aps,prd,showpacs,preprintnumbers,twocolumn]{revtex4}
\usepackage{epsfig}
\usepackage{latexsym,amsmath,amssymb,amstext}
\begin{document}
\title{Thermodynamics of the ideal overlap quarks on the lattice}
\author{Debasish\ \surname{Banerjee}}
\email{debasish@theory.tifr.res.in}
\affiliation{Department of Theoretical Physics, Tata Institute of Fundamental
         Research,\\ Homi Bhabha Road, Mumbai 400005, India.}
\author{R.\ V.\ \surname{Gavai}}
\email{gavai@tifr.res.in}
\affiliation{Department of Theoretical Physics, Tata Institute of Fundamental
         Research,\\ Homi Bhabha Road, Mumbai 400005, India.}
\author{Sayantan\ \surname{Sharma}}
\email{ssharma@theory.tifr.res.in}
\affiliation{Department of Theoretical Physics, Tata Institute of Fundamental
         Research,\\ Homi Bhabha Road, Mumbai 400005, India.}

\begin{abstract}
\end{abstract}
\preprint{TIFR/TH/08-10}
\begin{abstract}
The thermodynamics of massless ideal gas of overlap quarks has been
investigated both analytically and numerically for both zero and nonzero baryon
chemical potential. Any $\mu^2$-divergence is shown analytically to be absent
for a class of actions with nonzero chemical potential. All such actions are 
shown to violate chiral invariance. While the parameter
$M$ can be shown to be irrelevant in the continuum limit, as expected, it is
shown numerically that the continuum limit can be reached with relatively
coarser lattices for $1.5\leq M \leq1.6$.  Numerical limitations of the
existing method of introduction of chemical potential are demonstrated.
Finally we also show that the energy density for the massive overlap fermions
has the correct continuum limit.
\end{abstract}

\maketitle
\section{ Introduction }
Lattice QCD has so far provided the most reliable theoretical predictions for
the thermodynamics of quarks and gluons important for the ongoing experiments
at the Relativistic Heavy Ion Collider (RHIC), and may continue to do so for
those at the upcoming Large Hadron Collider(LHC).  While the equation of state
seems\cite{bieos} to exhibit robust features as one changes the number of light
quarks, $N_f$, the order of the phase transition and the transition temperature
$T_c$ seems\cite{LaePhi} to depend on it crucially.  Indeed, the location, and
even the existence of the critical point in the $\mu_B-T$ phase diagram is
expected\cite{RajWil} to depend on $N_f$, as a result of this dependence of the
order of the transitions on $N_f$.  Since the transition seems to be associated
with the restoration of the spontaneously broken chiral symmetry at high
temperatures, it is very important to study it using fermions having exact
chiral symmetries on the lattice.  The popular choices of the fermions employed
in simulations so far have either no chiral symmetry (Wilson fermions) on the
lattice or only partial chiral symmetry (staggered fermions).  For the latter
even $N_f$ is not well defined on the lattice and is typically taken to be the
anticipated continuum value.  Of course, these issues are expected to become
irrelevant in the continuum limit of vanishing lattice spacing, i.e, in the
limit when the number of sites in the temporal  direction becomes very large : $N_T
\to \infty$.  But they are likely to affect the current bunch of results
obtained on lattices up to $N_T = 8$.

In view of the experimental relevance of these issues, it would clearly be ideal
to employ fermions with exact chiral symmetry on lattice for investigations of
the QCD thermodynamics.  As is well-known by now, the overlap
fermions\cite{NeuNar} have such good chiral properties even on the lattice. The
corresponding fermion operator respects chiral symmetry at the expense of being
highly non-local, making the corresponding computations rather expensive.
Advances in both algorithms and the computer hardware may have brought such
investigations closer to reality today.  In this paper we investigate the
thermodynamics of the free overlap fermions with an aim to examine its continuum
limit both analytically and numerically.  For the above mentioned practical
reasons, we investigate numerically whether the irrelevant parameter $M$ (see
below for explicit definition) can be tuned optimally to recover the continuum
results on the smallest possible lattice size. These
predictions can be used in full QCD simulations with such fermions to do the
finite temperature calculations faster. A way to introduction of chemical
potential in the overlap formalism was proposed\cite{wet} and has been recently
studied\cite{gatt} numerically with $M=1$. It was shown that the known
canonical $\mu^2$ divergence at zero temperature did not appear.  Our analytical
work shows the absence of the divergence for all allowed $M$, $0 < M < 2$,
for the same class\cite{gav} of actions as the staggered fermions.

The plan of our paper is as follows. Section II deals with zero chemical
potential case.  The analytical ($N_T \to \infty$) results for the energy
density (equivalently pressure) at nonzero temperatures are derived and
the numerical result for finite $N_T$ are presented.  Corresponding results for
nonzero chemical potential are given in the next section, where the quark number
susceptibility are also presented in addition.  Section IV covers the case
of massive quarks.  A summary is provided in the final section V.

\section{Zero chemical potential}
The overlap Dirac operator\cite{NeuNar} has the following form  for massless 
fermions on asymmetric lattice with spacing $a$ and $a_4$ in the spatial and
temporal directions:
\begin{equation}
 D_{ov}=1+\gamma_5 sgn(\gamma_5 D_W)~,~
\end{equation}
where sgn denotes the sign function and
\begin{eqnarray}
\label{eqn:Dwil}
\nonumber
&&D_W(x,y) = (3+\frac{a}{a_4}-M)\delta_{x,y}
\\ \nonumber
&-&\frac{a}{a_4}[U^{\dagger}_{4}(x-\hat{4})
\delta_{x-\hat{4},y}\frac{1+\gamma_{4}}{2} 
+\frac{1-\gamma_{4}}{2}U_{4}(x)\delta_{x+\hat{4},y}]\\
&-&\sum_{i=1}^{3}
[U^{\dagger}_{i}(x-\hat{i})\delta_{x-\hat{i},y}\frac{1+\gamma_{i}}{2} 
+\frac{1-\gamma_{i}}{2}U_{i}(x)\delta_{x+\hat{i},y}]
\end{eqnarray}
is the standard Wilson-Dirac operator on the lattice but with a negative mass
term $M\in(0,2)$.  The overlap operator satisfies the Ginsparg-Wilson
relation\cite{wil} and has exact chiral symmetry on the lattice.  The
corresponding infinitesimal chiral transformations\cite{Lues} for $N_f =1 $ are 
\begin{equation}
\label{eqn:chrl}
\delta \psi = \alpha  \gamma_5(1 - \frac{1}{2}D_{ov}) \psi  ~~~{\rm and} ~~~  
\delta \bar \psi = \alpha \bar \psi (1 - \frac{1}{2}D_{ov})\gamma_5 ~,~
\end{equation}
$\psi$ and $ \bar \psi$ are the usual four component fermion and antifermion
fields.  They acquire a flavour index for higher $N_f$, with corresponding
modification in the transformations above similar to that in the continuum.
The expression for energy density and pressure can be obtained from the
partition function, $Z = {\rm det}~D_{ov}$, obtained by integrating the
quark-antiquark fields :
\begin{eqnarray}
\nonumber
 \epsilon &=& \frac{T^2}{V}
     \left.\frac{\partial\ln Z(V,T)}{\partial T}\right|_V, ~~~{\rm and}\\
 P &=& T
     \left.\frac{\partial\ln Z(V,T)}{\partial V}\right|_T, 
\end{eqnarray}
where the spatial volume $V = N^3 a^3$ and the temperature $T = (N_T a_4)^{-1}$
for an $N^3 \times N_T$ lattice.  We restrict  ourselves to $U=1$ here to focus
on the ideal gas limit.  Noting that the sign function for a matrix is
defined in terms of its eigenvalues, the energy density can be written as
\begin{eqnarray}
\label{eqn:x2}
\nonumber
\epsilon &=& -\frac{1}{N^3 a^3 N_T}\left(\frac{\partial ln (\prod_{_n}
\lambda_{n})}{\partial a_4}\right)_{a}\\
&=&-\frac{2}{N^3 a^3 N_T}\sum_{ \lambda_\pm }\left(\frac{\partial
   ln \lambda_{\pm}}{\partial a_4}\right)_{a}~,~
\end{eqnarray}
where the chiral nature of the eigenvalue spectrum in the free case was used in
the last line.  The eigenvalues of the free overlap operator in the momentum
space can be easily worked\cite{wb1, gatt} out to be
\begin{equation}
\label{eqn:lam}
 \lambda_{\pm}=1-\frac{sgn\left(\sqrt{h^2+h_5^2}\right) h_5 \pm i \sqrt{h^2}}{\sqrt{h^2+h_5^2}}~,~
\end{equation}  
where the variables h above are given by 

\begin{eqnarray}
\label{eqn:h15}
\nonumber
h_5&=&M-\sum _{j=1}^{3}(1-\cos(a p_j))-\frac{a}{a_4}(1-\cos(a_4 p_4))\\
\nonumber
h_j&=&-\sin(a p_j) \qquad \text{where}~j = 1,2,3\\
\nonumber
h_4&=&-\frac{a}{a_4}\sin(a_4 p_4)\\
h^2&=&h_1^2+h_2^2+h_3^2+h_4^2~.~
\end{eqnarray}
From the (anti)periodic fermion boundary conditions in the (time) space 
directions, the discrete $p_\mu$'s appearing in the equations above are seen
to have the following allowed values :
\begin{eqnarray}
\nonumber
ap_j &=& \frac{2n_j\pi}{N},n_j=0,..,(N-1), j=1,2,3 ~{\rm and}\\
ap_4 &=& \frac{(2n+1)\pi}{N_T},n=0,..,(N_T-1)
\end{eqnarray}
Note that the variables $h_i$ are all real.  Further, a simple algebra shows
that $ (h^2 + h_5^2 ) > 0$ for all ranges of interest for $M$, $a$ and $a_4$.
Since the sign term in eq.(\ref{eqn:lam}) is thus a constant, it does not
contribute to the derivative in eq.(\ref{eqn:x2}): it merely provides the
overall sign for the energy density.  Choosing $sgn(\sqrt{h^2+h_5^2})=1 $, the 
energy density becomes

\begin{eqnarray}
\label{eqn:x3}
\nonumber
\epsilon &=& \frac{2}{N^3 a^3 N_T} \sum_{p_j,p_4} \frac{ \frac{\partial h_5}{\partial a_4}-\frac{h_5}{h^2+h_5^2}\left(h_4 \frac{\partial h_4}{\partial a_4}+h_5 \frac{\partial h_5}{\partial a_4}\right)}{\sqrt{h^2+h_5^2}-h_5} \\
\nonumber
&=& \frac{2}{N^3 a^3 N_T} \sum_{p_j,p_4}\frac{h^2 \frac{\partial h_5}{\partial a_4}-h_5 h_4 \frac{\partial h_4}{\partial a_4}}{h^2 (h^2+h_5^2)}(\sqrt{h^2+h_5^2}+h_5)~,~ \\
\end{eqnarray}
where the summations are over all the discrete sets of momenta on the lattice.
The derivatives in the expression above are seen to be
\begin{eqnarray}
\frac{\partial h_4}{\partial a_4} &=& -\frac{h_4}{a_4}\\
\frac{\partial h_5}{\partial a_4} &=& \frac{a}{a_4^2}(1-\cos(a_4 p_4))
\end{eqnarray}
Similarly pressure $P$ can be computed by taking partial derivative with
respect to $a$, holding $a_4$ constant to obtain
\begin{equation}
\label{eqn:x3P}
P = \frac{-2}{3 N^3 a^2 a_4 N_T} \sum_{p_j,p_4}\frac{h^2 \frac{\partial h_5}{\partial a}-h_5 h_4 \frac{\partial h_4}{\partial a}}{h^2 (h^2+h_5^2)}(\sqrt{h^2+h_5^2}+h_5)~.~ \\
\end{equation}
The derivatives in the expression for pressure are 
\begin{eqnarray}
\frac{\partial h_4}{\partial a} &=&  \frac{h_4}{a}\\
\frac{\partial h_5}{\partial a} &=& - \frac{1}{a_4}(1-\cos(a_4 p_4))
\end{eqnarray}
Substituting the derivatives in eq.(\ref{eqn:x3}) and eq.(\ref{eqn:x3P}), one
finds the expected ideal gas equation of state $\epsilon = 3 P$, valid for
all values of $a$ and $a_4$.   We shall therefore focus in the remainder only 
on the  energy density for free overlap quarks on the lattice and evaluate it
by setting $a_4=a$.   We introduce a more compact notation for doing so: 
\begin{eqnarray}
\nonumber
 h_5&=&g+\cos\omega\\
\nonumber
h^2&=&f+\sin^2\omega\\
h^2+h_5^2&=&d+2g \cos\omega~,~
\end{eqnarray}
where $\omega = ap_4$ and the functions $g$, $f$ and $d$ are given by
\begin{eqnarray}
\label{eqn:aux}
\nonumber
g&=&M-4+b, ~{\rm with}\\
\nonumber
b&=&\cos(a p_1)+\cos(a p_2)+\cos(a p_3)\\
\nonumber
f&=& h_1^2+h_2^2+h_3^2\\
\nonumber
d&=& 4+(M-4)^2+2(M-4) b +c, ~{\rm with} \\
c &=& \sum_{i<j< 4} 2 \cos(a p_i) \cos(a p_j)~.~
\end{eqnarray}
It may be noted that the $g$, $d$ and $f$ depend only on spatial momenta $p_j$
and enable us to write down the $ap_4 = \omega$-dependence of the energy 
density explicitly:

\begin{eqnarray}
\label{eqn:xxx}
\epsilon a^4& =& \frac{2}{N^3 N_T}\sum_{p_j,n} \left[(g+ \cos\omega_n)+ \sqrt{d+2 g \cos\omega_n}\right] \nonumber\\
&\times& \left[ \frac{(1-\cos\omega_n) }{d+2g \cos\omega_n} + \frac{\sin^2\omega_n(g+\cos\omega_n)}{(d+2g \cos\omega_n)(f+\sin^2\omega_n)} \right]~.~ \nonumber\\
\end{eqnarray}

As shown in the Appendix \ref{zeroMU}, the summation over the Matsubara 
frequencies can be carried out using the standard contour integral techniques,
resulting in the energy density on the lattice to be,
\begin{eqnarray}
\label{eqn:lat1}
\nonumber
\epsilon a^4& =&\frac{4}{N^3} \sum_{p_j}\left [\frac{\sqrt{f}}{\sqrt{1+f}}\right]\frac{1}{e^{ N_T\sinh^{-1}\sqrt{f}}+1}\\
&+&\epsilon_3+\epsilon_4~,~
\end{eqnarray}
where $\epsilon_3$, $\epsilon_4$ terms come from from the line integrals 3 and
4 in Figure \ref{cont} respectively. Their explicit $N_T$ dependence indicates
that they contribute to the energy density on the lattice.  However, they do
not do so in the continuum limit, as we shall see below.

In order to take the continuum limit of $a\rightarrow0$, we let 
$N,N_T\rightarrow \infty$ such that T and $VT^3$ is kept constant. Each
summation over momenta  is replaced by an integral in this limit:
\begin{equation}
\label{eqn:elat}
\frac{1}{N}\sum_{p_j} \rightarrow \frac{a}{2 \pi} \int_{-\infty}^{\infty} d p_j.
\end{equation}
Further the integration variable $\omega = ap_4$ can be traded for $p_4$,
pushing the the branch points at $\pm \pi \pm i \text{cosh}^{-1}\frac{d}{2g}$
to infinity faster than the contours 3 and 4 are pushed. The line integrals.
and hence the terms $\epsilon_3$ and $\epsilon_4$ vanish. Since the poles at $i~
\text{sinh}^{-1}\sqrt{f}$ scale as $a$ in this limit, they continue to be
enclosed in the contour at a finite $p_4$ and do contribute to the energy
density. This can, of course, be explicitly checked algebraically by taking
the limit of the eq.(\ref{eqn:lat1}) to obtain the expression for the 
continuum energy density as 
\begin{eqnarray}
\label{eqn:sb}
 \nonumber
 \epsilon_{SB} &=&\frac{2}{(2 \pi)^3 } \left(2\int \prod_{j=1}^{3}dp_j\frac{E}{1+e^{E/T}}\right)\\
&=&\frac{7 \pi^2}{60}T^4 ~,~
\end{eqnarray}
where $E=\sqrt{p_{1}^2+p_{2}^2+p_{3}^2}$ is the energy of the massless quarks.

\subsection{Numerical evaluation}

In this subsection we investigate the lattice energy density of the
eq.(\ref{eqn:xxx}) numerically by summing over all the momenta.  Our aim is i)
to estimate the importance of the terms $\epsilon_3$ and $\epsilon_4$ in it on
lattices of practical sizes, and ii) to find out the role $M$ plays on finite
lattices.  In particular, it would be good to know if there exists a range of
the irrelevant parameter $M$ for which the energy density converges to that of
the continuum ideal Fermi gas on reasonable, i.e. computationally inexpensive,
lattice sizes.  Since we have shown the existence of the continuum limit for
the entire allowed range of $M$ in the previous subsection, it is clear that a
sufficiently fine lattice must eventually yield the correct result for any $M$.  

\begin{figure}
\begin{center}
 \includegraphics[scale=0.7]{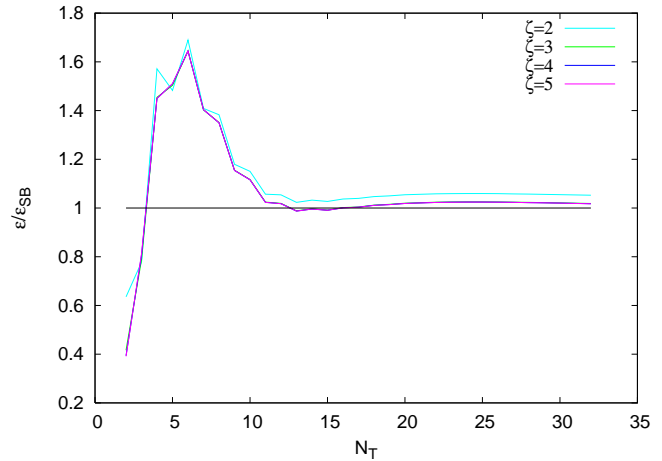}
\caption{The variation of the ratio $\epsilon /\epsilon_{SB} $  with 
$N_T$ for $M=1.55$ and $\zeta = 2-5$.}
\label{zeta}
\end{center}
\end{figure}

\begin{figure}
\begin{center}
\includegraphics[scale=0.7]{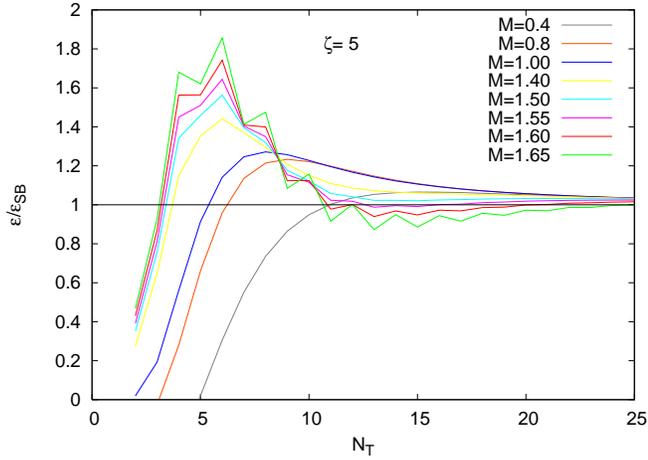}
\caption{The variation of the ratio $\epsilon /\epsilon_{SB} $  with 
$N_T$ for different M and $\zeta = 5$.}
\label{Mall}
\end{center}
\end{figure}

\begin{figure}
\begin{center}
 \includegraphics[scale=0.37]{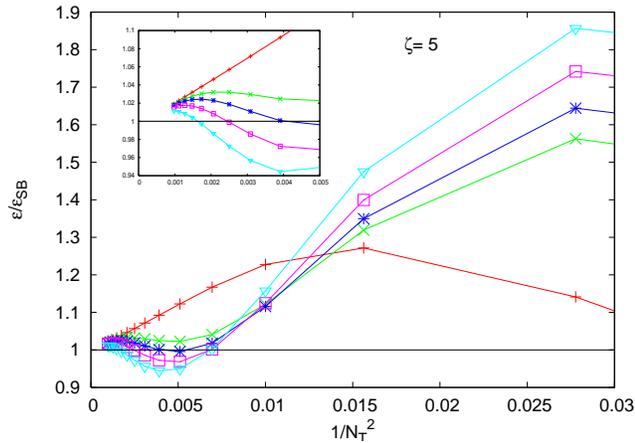}
\caption{The variation of the ratio $\epsilon /\epsilon_{SB} $  with 
$1/N_T^2$ for different $M \geq1$ and $\zeta = 5$. The plusses, crosses, stars, 
boxes and triangles denote M=1.0, 1.50, 1.55, 1.60 and 1.65 respectively.}
\label{nt2}
\end{center}
\end{figure}

In general, the dimensionless lattice energy density will be of the form,
\begin{equation}
\label{eqn:laten}
 E=A(M)+\frac{B}{N_T^4}+\frac{C(M)}{N_T^6}+\frac{D(M)}{N_T^8}+...
\end{equation}
where the coefficients $A(M)$ and $B$ are the usual vacuum and the $T^4$
contributions, while $C(M)$ and $D(M)$ are finite lattice spacing artifacts.
For each value of M and aspect ratio, defined as $\zeta=N/N_T$, the energy
density on the lattice was calculated as a function of the $N_T$.  Clearly $A$
is the dominant contribution and its removal turned out to be a tricky issue
governed by the precision of our computations.   Fitting the above form to
obtain $C$ or $D$ was therefore not feasible.  The zero temperature part of the
energy density was calculated from eq.(\ref{eqn:xxx}) by taking
$N_T\rightarrow\infty$ and  a large spatial extent, keeping the lattice spacing
finite. The resulting integral over $\omega$ was done numerically for each M to
estimate the zero temperature contribution. Subtracting the zero temperature
part from the energy density and dividing the resultant $\epsilon$ by
$\epsilon_{SB}$ gives us a ratio which we employ for further studies.  Figure
\ref{zeta}  displays the ratio $\epsilon /\epsilon_{SB}$ as a function of $N_T$
for $M=1.55$ and various aspect ratios $\zeta$.  A mild dependence on $\zeta$
is visible for lower values but in each case the curve approaches to unity by
$N_T = 12$, signalling the onset of continuum limit.  Figure \ref{Mall} exhibit
the $M$-dependence of the same ratio for a fixed $\zeta=5$ for the range $0.4
\le M \le 1.65$.   A range of $1.5 \le M \le 1.6$ emerges as the favoured one
because all the M-dependent terms are seen to be minimum there and hence the
continuum limit is reached faster.  On smaller lattices with $N_T = 4-8$, the
lattice results are seen to be 1.6-1.8 times larger in this range of $M$.  For
other values of $M$, the continuum limit is seen to be approached slowly; even
an $N_T=25$ seems not enough.  For larger $M$, we also observed oscillations as
$N_T$ changed between odd and even, limiting our effort to increase the
$M$-range further. The values of $\epsilon/\epsilon_{SB}$ for $N_T=4-16$ and
different M are tabulated in Tables \ref{ez2}, \ref{ez5} for easy reference.
We note that the $\zeta=3,4$ results are the same as that for $\zeta=5$ as seen
from Figure \ref{zeta}.  In order to estimate the size of the $1/N_T^2$
correction term for different values of M, the same ratio is plotted as a
function of $1/N_T^2$ in Figure \ref{nt2}.  From the plot, it is evident that
the correction terms die down very fast for $ 1.50 \leq M\leq1.60$ and the
continuum limit is reached  within 2-3 \% already for $N_T=12$ whereas for
$M=1$ they are relevant even for $N_T\geq12$.  Of course, the continuum
extrapolation for $M=1.0$ is easier than for $1.50\leq M \leq1.60$  due to the
nonlinearities present for the latter.  Note, however, that energy density for
at least three different lattice sizes with $N_T=10, 12, 14$ need to be
computed for such an extrapolation.  On the other hand, although the
extrapolation for $1.50\leq M\leq 1.60$ is difficult due to the complex
variation seen in Figure \ref{nt2}, the deviation from the continuum value is
within the typical accuracy range of the current lattice results, making it an
optimal range for simulations.  It should also be noted that the corrections
for the overlap fermions for $M\sim1.55$ for $N_T < 12 $ are smaller than
compared to the Wilson and the staggered case\cite{hegde} as well.  Ref.
\cite{hegde} deals with $p/p_{SB}$ which we showed above to be identical to the
$\epsilon/\epsilon_{SB}$ for the overlap ideal gas.
\begin{table}
\caption{$\epsilon/\epsilon_{SB}$ values for diff. M for $\zeta=2$}
\begin{tabular}{@{\extracolsep{\fill}}|c||c|c|c|c|c|}
\hline
$N_T$&M=1.0&1.50&1.55&1.60&1.65\\
\hline
4&0.630 &1.453  & 1.571 &1.697 & 1.828 \\
6&1.194 &1.606 &1.690  &1.792  &  1.914 \\
8&1.316&1.355 &1.383  & 1.431 & 1.506\\
10&1.268 &1.158 &1.150 &1.156  &  1.186\\
12&1.206 & 1.078 &1.054 &1.036 & 1.033\ \\
14&1.160 &1.060 &1.032 &1.004 & 0.983\\
16&1.129 &1.061  &1.037  & 1.008  & 0.979\\
\hline
\end{tabular}
\label{ez2}
\end{table}

\begin{table}
\caption{$\epsilon/\epsilon_{SB}$ values for diff. M for $\zeta=5$}
\begin{tabular}{@{\extracolsep{\fill}}|c||c|c|c|c|c|}
\hline
$N_T$&M=1.0&1.50&1.55&1.60&1.65\\
\hline
4& 0.561& 1.342& 1.450& 1.563& 1.681\\
6& 1.141& 1.563& 1.644& 1.742& 1.857 \\
8& 1.272 & 1.319& 1.350& 1.399& 1.475\\
10& 1.228& 1.122& 1.116& 1.124& 1.157\\
12& 1.167& 1.041& 1.018& 1.001& 1.002\\
14& 1.123& 1.023& 0.996& 0.969& 0.950\\
16& 1.092& 1.025& 1.001& 0.972& 0.944\\
\hline
\end{tabular}
\label{ez5}
\end{table}

\begin{figure}
\begin{center}
\includegraphics[scale=0.7]{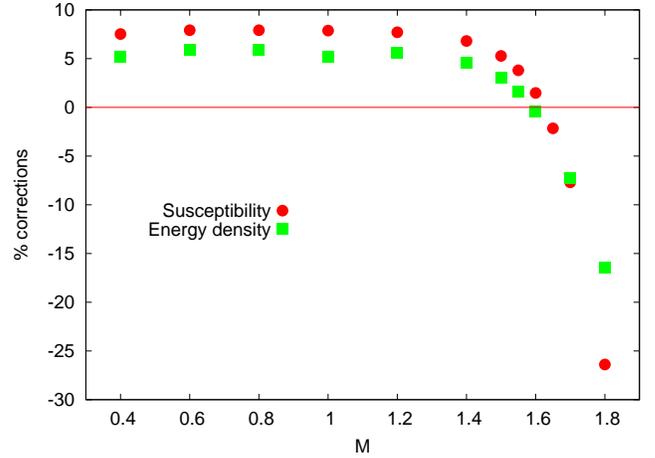}
\caption{The estimated finite lattice spacing corrections for the energy 
density and susceptibility at $\mu=0$ in percentage as a function of $M$.}
\label{sbdv}
\end{center}
\end{figure}

Filled squares in Figure \ref{sbdv} show the percentage average deviations of 
the ratio from unity due to lattice artifact terms as a function of $M$ for large $N_T$
values($N_T\geq 18$).  It shows marginal dependence on $M$ for $M < 1.2$ 
but the deviation itself is about 5-6 \%.  For larger $M$, the data show a dip, 
indicating clearly that the thermodynamics of free fermions favours the 
optimum value of M to lie between 1.50-1.60, with a deviation of only about 2.5 \% or lower.
 
Comparing our results with other studies of thermodynamics of free fermions done
with improved actions\cite{wiese} and also with overlap fermions ($M=1$) in
2-D\cite{hip} as well as in 4-D\cite{gatt,hegde}, we find that i) there are
larger deviations in higher dimensions and ii) the oft-favoured choice of $M=1$
favours rather poorly on finite lattices.  Indeed, one can significantly reduce
the corrections to the energy density of overlap fermions due to the lattice
artifacts with a proper choice of M.


\section{Nonzero chemical potential}
The chemical potential is usually introduced as the Lagrange multiplier to
investigate thermodynamics at constant conserved number. Constructing the
relevant number operator for the overlap Dirac fermions is not easy due to its
nonlocality\cite{kiku} and may even be not unique\cite{mand}. Instead of
deriving the conserved number, one may make an inspired guess for it such that
it has the right continuum limit.  One such proposal for introducing the
chemical potential for the overlap operator is \cite{wet} to introduce it in the
$D_W$ as one would for the usual Wilson fermions:
\begin{equation}
\label{eqn:ovwet}
 D_{ov}=1+\gamma^5 sgn(\gamma^5 D_W(\hat \mu))~,~
\end{equation}
where the chemical potential $\hat{\mu} = \mu a_4$ appears only as multiplying
factors $\exp(\hat{\mu})$ and $\exp(-\hat{\mu})$ to the links $U_4$ and
$U_4^\dag$ respectively in eq.(\ref{eqn:Dwil}). This, of course, renders
$\gamma_5 D_W(\mu)$ to be non-hermitian, necessitating an extension of the usual
definition of the sign function.  The natural choice \cite{wet} was to use the
sign of the real part of the eigenvalues of $\gamma_5 D_W(\mu)$ in the equation
above.  It is important to note that the extended  sign function it is not
defined for purely imaginary eigenvalues.  Numerical simulations were
performed\cite{gatt} for an ideal gas of overlap fermions to show that the above
way of introducing $\mu$ does not encounter any quadratic divergences at zero
temperature.  Such divergences were known to arise\cite{karsch,gavbil} for
staggered and Wilson fermions, if $\mu$ was introduced naively as a coefficient
of the conserved number. These were eliminated by the the choice of the
$\exp(\pm \hat{\mu})$ factors.  A more general way to introduce the chemical
potential is, of course, to introduce functions $K(\hat{\mu})$ and
$L(\hat{\mu})$ in place of the factors $\exp(\hat{\mu})$ and $\exp(-\hat{\mu})$
respectively such that $K(\hat{\mu}) = 1 + \hat{\mu} + {\cal O}(\hat{\mu}^2) $
and $L(\hat{\mu}) = 1 - \hat{\mu} + {\cal O}(\hat{\mu}^2) $, It was shown
\cite{gav} that the quadratic divergences are avoided if $K(\hat{\mu}) \cdot
L(\hat{\mu}) = 1$. 

Here we follow that idea and introduce chemical potential in the overlap Dirac
operator through the $K$ and $L$ factors in $D_W$ and study where the condition
to eliminate the quadratic divergences remains the same.  Introducing
\begin{eqnarray}
\nonumber
 \frac{K(\hat{\mu})-L(\hat{\mu})}{2}&=&R\sinh\theta\\
\frac{K(\hat{\mu})+L(\hat{\mu})}{2}&=&R\cosh\theta ~,~
\end{eqnarray}
one can follow through the steps of the previous section to find that 
the free overlap operator in the momentum space can again be written in 
terms of the $h_i$ of eq.(\ref{eqn:h15}) but with $h_4$ and $h_5$ changed to :
\begin{eqnarray}
\label{eqn:h15mu}
\nonumber
h_5&=&M-\sum _{j=1}^{3}(1-\cos(a p_j))-\frac{a}{a_4}(1-R\cos(a_4 p_4 -i \theta))\\
h_4&=&-\frac{a}{a_4}R\sin(a_4 p_4 -i \theta)~.~
\end{eqnarray}

The energy density in presence of finite chemical potential is defined as
\begin{eqnarray}
\label{eqn:cp}
\epsilon(\mu) &=& -\frac{1}{N^3 a^3 N_T}\left(\frac{\partial ln \det D}{\partial
  a_4}\right)_{a, a_4 \mu}\\
&=&-\frac{2}{N^3 a^3 N_T}\left(\frac{\partial ln (\lambda_+ \lambda_-)}{\partial
  a_4}\right)_{a, a_4 \mu}
\end{eqnarray}
In the following we assume that the sign function is always defined and is 
$+1$, as for the $\mu =0$ case.  We shall comment on this assumption later. The
energy density is obtained using  eq.(\ref{eqn:cp}) and setting $a=a_4$.
\begin{eqnarray}
\label{eqn:cp1}
\nonumber
\epsilon a^4& =& \frac{2}{N^3 N_T}\sum_{p_j,n}\left[ \frac{1-R\cos(\omega_n-i\theta) }{d_R+2gR \cos(\omega_n-i\theta)}\right.\\
\nonumber
&+&\left.\frac{R^2\sin^2(\omega_n-i\theta)(g+R\cos(\omega_n-i\theta))}{(d_R+2gR \cos(\omega_n-i\theta))(f+R^2\sin^2(\omega_n-i\theta))}\right] \\
\nonumber
&\times&\left[g+ R\cos(\omega_n-i\theta)+ \sqrt{d_R+2 g R\cos(\omega_n-i\theta)}\right]\\
\end{eqnarray}
Note that the summand in the equation above has the same functional form as
that in eq.(\ref{eqn:xxx}).  Indeed the only changes are: $d_R=f+g^2+R^2$ replaces
d of eq.(\ref{eqn:aux}), $\omega \to\omega -i \theta$ and the factor $R$ multiplies 
each sine/cosine term. Let
us therefore denote the summand in eq.(\ref{eqn:cp1}) as $F(R, \omega 
-i \theta)$. Comparing eq.(\ref{eqn:h15mu}) with eq.(\ref{eqn:h15}), and 
using the expression for the pressure given in eq.(\ref{eqn:x3P}), one
again finds that the equation of state $\epsilon=3P$ also holds in the presence
of a chemical potential, $\mu$.  An additional new physical observable that can
be computed is the number density, defined as,
\begin{equation}
n = \frac{1}{N^3 a^3  N_T}\left(\frac{\partial ln \det D}{\partial
\hat{\mu}  }\right)_{a_4}
\end{equation}
In terms of h's the previous expression can be calculated explicitly,
\begin{eqnarray}
\label{eqn:np1}
\nonumber
&&n a^3 = \frac{-2i}{N^3 N_T}\sum_{p_j,n}\left[ R\sin\left(\omega_n-i\theta\right)\right.\\
\nonumber
&\times&\left(\frac{gR\cos(\omega_n-i\theta)+R^2+f}{(d_R+2gR \cos(\omega_n-i\theta))(f+R^2\sin^2(\omega_n-i\theta))}\right) \\
\nonumber
&\times&\left.\left(g+ R\cos(\omega_n-i\theta)+ \sqrt{d_R+2 g R\cos(\omega_n-i\theta)}\right)\right]\\
&=&\frac{-2i}{N^3 N_T}\sum_{p_j,n}F_N(R,\omega_n-i\theta)~.~
\end{eqnarray}

\subsection{Analytic Results}
In order to obtain the condition for removing the divergences,
we first calculate the energy density at zero temperature i.e for the limit
$N_T\rightarrow\infty$ at finite a. The frequency sum 
$\frac{1}{N_T}\Sigma$ in eq.(\ref{eqn:cp1}) gets replaced by the integral
$\frac{1}{2\pi}\int_{-\pi}^{\pi}d\omega$. Subtracting the vacuum contribution
corresponding to $\mu = 0$, i.e., $R = 1, \theta =0 $, the energy density at 
zero temperature is given by
\begin{equation}
 \epsilon a^4=\frac{1}{\pi N^3 }\sum_{p_j}\left[\int_{-\pi}^{\pi}F(R,\omega-i\theta)d\omega-\int_{-\pi}^{\pi}F(\omega)d\omega\right]
\end{equation}
\begin{figure}
\begin{center}
 \includegraphics[scale=0.3]{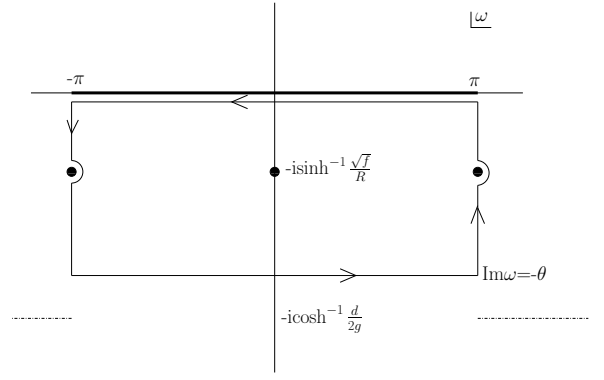}
\caption{Contour chosen for evaluating the energy density and the number
density for nonzero of chemical potential at zero temperature. The thick line
indicates the Matsubara frequencies  while the filled circles denote the poles
of $F(R, \omega)$.}
\label{zmuc}
\end{center}
\end{figure}
Choosing the contour shown in Figure \ref{zmuc}, the expression above can be
evaluated in the complex $\omega$-plane as
\begin{eqnarray}
\nonumber
 \epsilon a^4&=&\frac{1}{\pi N^3 }\sum_{p_j}\left[2\pi i\sum_{i} \text{Res}~F(R,\omega_i)\right.\\
\nonumber
&-&\int_{\pi-i\theta}^{\pi}F(R,\omega)d\omega-\int_{\pi}^{-\pi}F(R,\omega)d\omega\\
&-&\left.\int_{-\pi}^{-\pi-i\theta}F(R,\omega)d\omega-\int_{-\pi}^{\pi}F(\omega)d\omega\right]~.~
\end{eqnarray}
The second and fourth terms cancel since F is an even function which satisfies
$F(\pi+i\eta)=F(-\pi+i\eta)$. Hence, we obtain
\begin{eqnarray}
\label{eqn:emuz}
\nonumber
\epsilon a^4&=&\frac{1}{\pi N^3}\sum_{p_j}\left[2\pi R_3\Theta\left(\frac{K(\hat{\mu})-L(\hat{\mu})}{2}-\sqrt{f}\right)\right.\\
&+&\left.\int_{-\pi}^{\pi}F(R,\omega)d\omega-\int_{-\pi}^{\pi}F(\omega)d\omega\right]~,~
\end{eqnarray}
where -i$R_3$ is the residue of the function $F(R,\omega)$ at the pole
$-i$~sinh$^{-1}(\sqrt{f}/R)$ and is given by
\begin{eqnarray}
\nonumber
 R_3 &=& \frac{\sqrt{f}(g+\sqrt{f+R^2}+\sqrt{d_R+2g\sqrt{f+R^2}})}{2\sqrt{f+R^2}(d_R+2g\sqrt{f+R^2})} \\
     &\times& (g+\sqrt{f+R^2})~.~
\end{eqnarray}
Noting that the number density in eq.(\ref{eqn:np1}) has the same pole
structure as the energy density, with only the residues being different in the
two cases, the latter can also be calculated in the same way to obtain,
\begin{eqnarray}
\label{eqn:nomu}
\nonumber
n a^3&=&\frac{1}{\pi N^3}\sum_{p_j}\left[2\pi R_4\Theta\left(\frac{K(\hat{\mu})-L(\hat{\mu})}{2}-\sqrt{f}\right)\right.\\
&-i&\left.\int_{-\pi}^{\pi}F_N(R,\omega)d\omega+i\int_{-\pi}^{\pi}F_N(\omega)d\omega\right]~,~
\end{eqnarray}
where $R_4$ is the residue of the function $F_N(\omega)$ at the pole
$-i$~sinh$^{-1}(\sqrt{f}/R)$
given by, 
\begin{eqnarray*}
 \nonumber
 R_4 &=& \frac{(g+\sqrt{f+R^2}+\sqrt{d_R+2g\sqrt{f+R^2}})}{2\sqrt{f+R^2}(d_R+2g\sqrt{f+R^2})} \\
     &\times& (g\sqrt{R^2+f}+R^2+f)~.~
\end{eqnarray*}
It is clear from both eqs.(\ref{eqn:emuz}) and (\ref{eqn:nomu})
that the condition $R = 1$ cancels the two
integrals in each of them, yielding the canonical forms of the Fermi surface.
For $R\neq1$, there will in general be violations of the Fermi
surface on the lattice.  Moreover, in the continuum limit $a \to 0$, 
one will in general have the $\mu^2/a^2$-divergences for $R\neq1$ in both
the energy density and the number density.  The 
condition to obtain the correct continuum values of $\epsilon=\mu^4/4\pi^2$ 
and $n=\mu^3/3\pi^2$ can also be seen to be the expected 
$K(\hat{\mu})-L(\hat{\mu})=2\hat{\mu} +O(\hat{\mu}^2)$.  
Note that the earlier work\cite{gav} on staggered
fermions employed the exact number density on the lattice which is not the case
for the overlap fermions here.  That one obtains still identical conditions in
both the cases suggests that it is indeed the behaviour near the continuum limit
which dictates these conditions.

Another crucial difference is that the introduction of the functions $K$ and
$L$ for the staggered fermions still leaves the action invariant under the
chiral transformations due to the locality of the action.  This is true for the
full theory, i.e., even after the link variables , $U^{\mu}_x$ are restored. On
the contrary, one can easily check that one breaks the chiral invariance in the
case of the overlap fermions by these functions $K,L$, or $\exp(\pm
\hat{\mu})$.  As defined in eq.(\ref{eqn:chrl}), the chiral transformation
involves $D_{ov}(\hat{\mu}=0)$, while the action for $\mu \neq 0$ for the
overlap fermions has $D_{ov}(\hat{\mu})$ of the eq.(\ref{eqn:ovwet}).  By
construction, the latter does satisfy the Ginsparg-Wilson relation \cite{wil}
with the $\mu$-dependent overlap Dirac operator on both sides :
\begin{equation}
\label{eqn:gwmu}
\big\{ \gamma_5, D_{ov}(\hat{\mu}) \big\} = D_{ov}(\hat{\mu}) \gamma_5
D_{ov}(\hat{\mu}) ~.~
\end{equation}
Unfortunately though, it is not sufficient to guarantee invariance under
the chiral transformation in eq.(\ref{eqn:chrl}), as it does not have any
$\mu$-dependence.  Indeed, the variation of action under the chiral
transformation of eq.(\ref{eqn:chrl}) is
\begin{eqnarray}
\nonumber
\delta S &=& \alpha \sum_{x,y} \bar \psi_x \big[ \gamma_5 D_{ov}(\hat{\mu}) + 
D_{ov}(\hat{\mu}) \gamma_5 \\
&-& \frac{1}{2}  D_{ov}(0) \gamma_5 D_{ov}
(\hat{\mu}) -\frac{1}{2} D_{ov}(\hat{\mu}) \gamma_5 D_{ov}(0) \big]_{xy} 
\psi_y ~,~
\end{eqnarray}

which clearly does not vanish in spite of eq.(\ref{eqn:gwmu}).

Finally, using the same techniques to evaluate the Matsubara frequencies sum,
the energy density at non-zero temperature and chemical potential can be 
computed analytically. Further details are given in the Appendix 
\ref{nonzeroTMU}. The final expression is, 
\begin{eqnarray}
\label{elatmu}
\nonumber
 \epsilon a^4&=&\frac{2}{N^3}\sum_{p_j}\left[\frac{\sqrt f}{\sqrt {1+f}}\frac{1}{e^{(\sinh^{-1}\sqrt f-\hat{ \mu}) N_T}+1}\right.\\
\nonumber
&+&\frac{\sqrt f}{\sqrt {1+f}}\frac{1}{e^{(\sinh^{-1}\sqrt f+\hat{ \mu}) N_T}+1}\\
&+&\left. \epsilon_{3\mu}+\epsilon_{4\mu} \right]
\end{eqnarray}
The terms $\epsilon_{3\mu}$ and $\epsilon_{4\mu}$ are contributions of the line
integrals in Figures \ref{cont2} and \ref{cont3}.  They vanish in the continuum
limit leaving only the contribution due to the residues of the poles shown in
those figures.  It can be shown explicitly that the expression for the lattice
energy density reduces to the well known\cite{rothe} result in the continuum.
\begin{equation}
 \epsilon=\frac{2}{(2\pi)^3}\int \frac{E ~\prod_{j=1}^{3} dp_j}{1+e^{\frac{E+ \mu}{T}}}
+\frac{2}{(2\pi)^3}\int \frac{E ~\prod_{j=1}^{3} dp_j}{1+e^{\frac{E- \mu}{T}}}
\end{equation}

\subsection{Numerical evaluation}
As in section II.A, our aim of presenting numerical results by carrying out the
sums over all momenta in eq.(\ref{eqn:elatmu}) is to find out the importance of
lattice artifacts in form of the terms involving 
$\epsilon_{3\mu}$ and $\epsilon_{4\mu}$, resulting from the line integrals 3
and 4, and to look for the role of $M$.  The focus here is, of course, on the
chemical potential.  We therefore consider two observables here. One is the
change in the energy density, $\Delta \epsilon (\mu,T) = \epsilon(\mu,T) -
\epsilon(0,T)$.  In continuum it is given by
\begin{equation}
\label{eqn:dem}
\frac{\Delta \epsilon(\mu,T)}{T^4}=\frac{\mu^4}{4\pi^2
T^4}+\frac{\mu^2}{2T^2}~.~
\end{equation}
The other quantity we consider is the quark number susceptibility at 
$\hat{\mu=0}$. 
For the free overlap fermions, it is given for any $\hat{\mu}$ by
\begin{equation}
\chi=\frac{1}{N^3 a^2 N_T}\left(\frac{\partial^2 ln \det D}{\partial
\hat{\mu}^2  }\right)_{a_4}~,~
\end{equation}
which can be worked out to be
\begin{eqnarray}
 \nonumber
\chi&=&\frac{2i}{N_T N^3 a^2}\sum_{p_j,p_4}\left[\frac{-\left(h^2 h_4+h_4 h_5 \cos(ap_4-i\hat{\mu} )\right)u}{s^4(s-h_5)^2}\right.\\
&+&\left.\frac{ v}{s^2(s-h_5)}\right]~.~
\end{eqnarray}
u and v are in the expression above are
\begin{eqnarray}
\nonumber
 u&=&2(s-h_5)(h_4\frac{\partial h_4}{\partial \hat{\mu} }+h_5\frac{\partial h_5}{\partial\hat{\mu} })\\
&+&s^2(\frac{\partial s}{\partial \hat{\mu}}-\frac{\partial h_5}{\partial
\hat{\mu}})~,~
\end{eqnarray}
\begin{eqnarray}
\nonumber
 v&=&\frac{\partial h_4}{\partial \hat{\mu}}(2h_4^2+h^2+h_5\text{cos}(ap_4-i\hat{\mu}))\\
&+& h_4\frac{\partial h_5}{\partial \hat{\mu}}\cos(ap_4-i\hat{\mu})+i h_4
h_5\text{sin}(ap_4-i\hat{\mu})~,~
\end{eqnarray}
and
\begin{equation}
 s^2=h^2+h_5^2~.~
\end{equation}

Again in the continuum, the susceptibility is known to be
\begin{equation}
\label{eqn:sus}
\chi(\mu)=\frac{\mu^2}{\pi^2}+\frac{T^2}{3}~.~ 
\end{equation}

Our computations for the energy density were performed keeping the ratio
$r=\mu/T=\hat{\mu} N_T$ fixed, yielding a constant $\Delta \epsilon/T^4$ in the
continuum from eq.(\ref{eqn:dem}).  Our choices of $r$ were restricted by
the fact that on lattices with odd $N_T$, eigenvalues of $\gamma_5 
D_W(\hat \mu)$ can turn purely imaginary for sufficiently large $\hat \mu$.
This is related to the fact that $(\gamma_5 D_W)^\dag \gamma_5 D_W$ has 
$h^2+h_5^2$ as eigenvalues and 
\begin{eqnarray}
\nonumber
{\rm Re~} (h^2 +h_5^2) = g^2 +1+f + 2g \cos \omega {\rm~cosh} \hat \mu \\
{\rm Im~} (h^2 +h_5^2) =  2g \sin \omega {\rm~sinh} \hat \mu ~.~
\end {eqnarray}
has zero imaginary part at $\omega = \pi$ with negative real part for $\mu \ge
\mu_c$.  The sign function is undefined for such cases.  Indeed, in the
interacting case it may even be possible to get such purely imaginary argument
of the sign function of the overlap Dirac operator for all $N_T$. 

\begin{figure}
\begin{center}
 \includegraphics[scale=0.7]{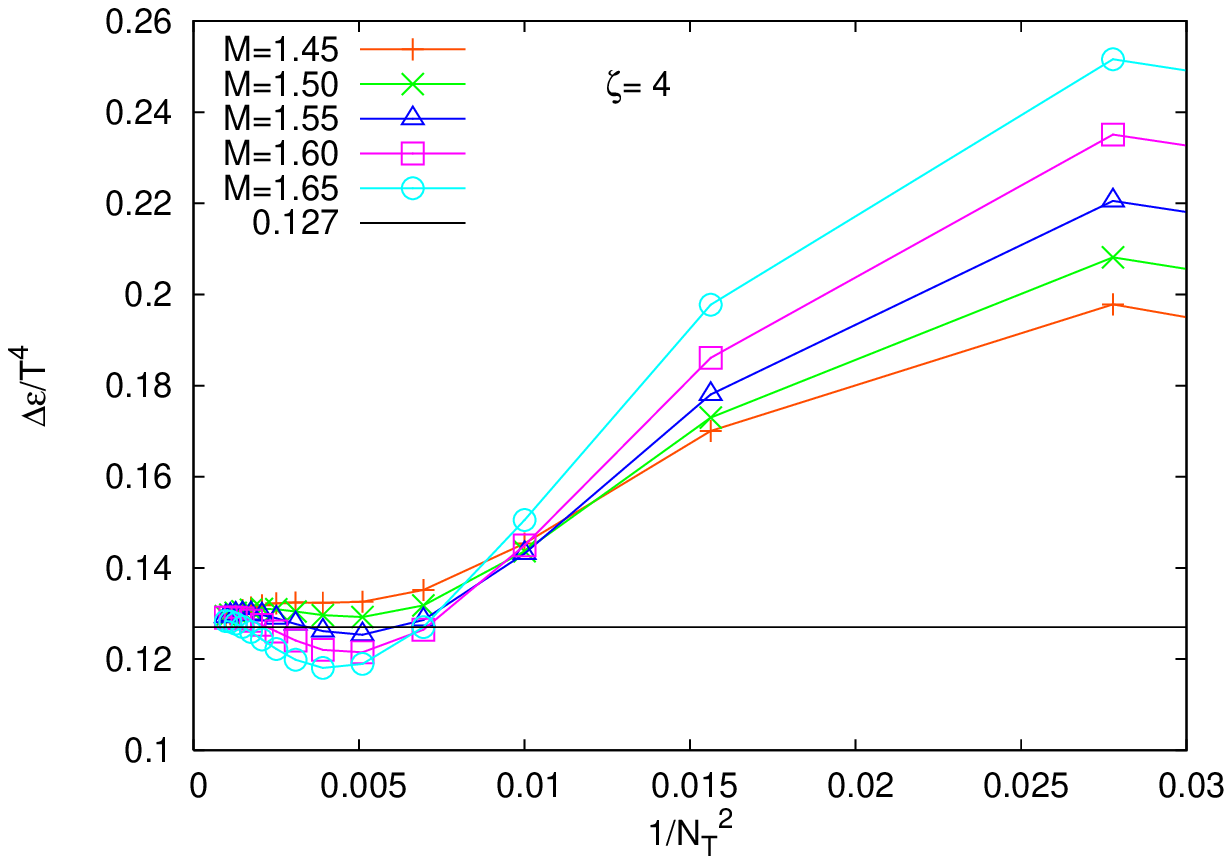}
 \includegraphics[scale=0.7]{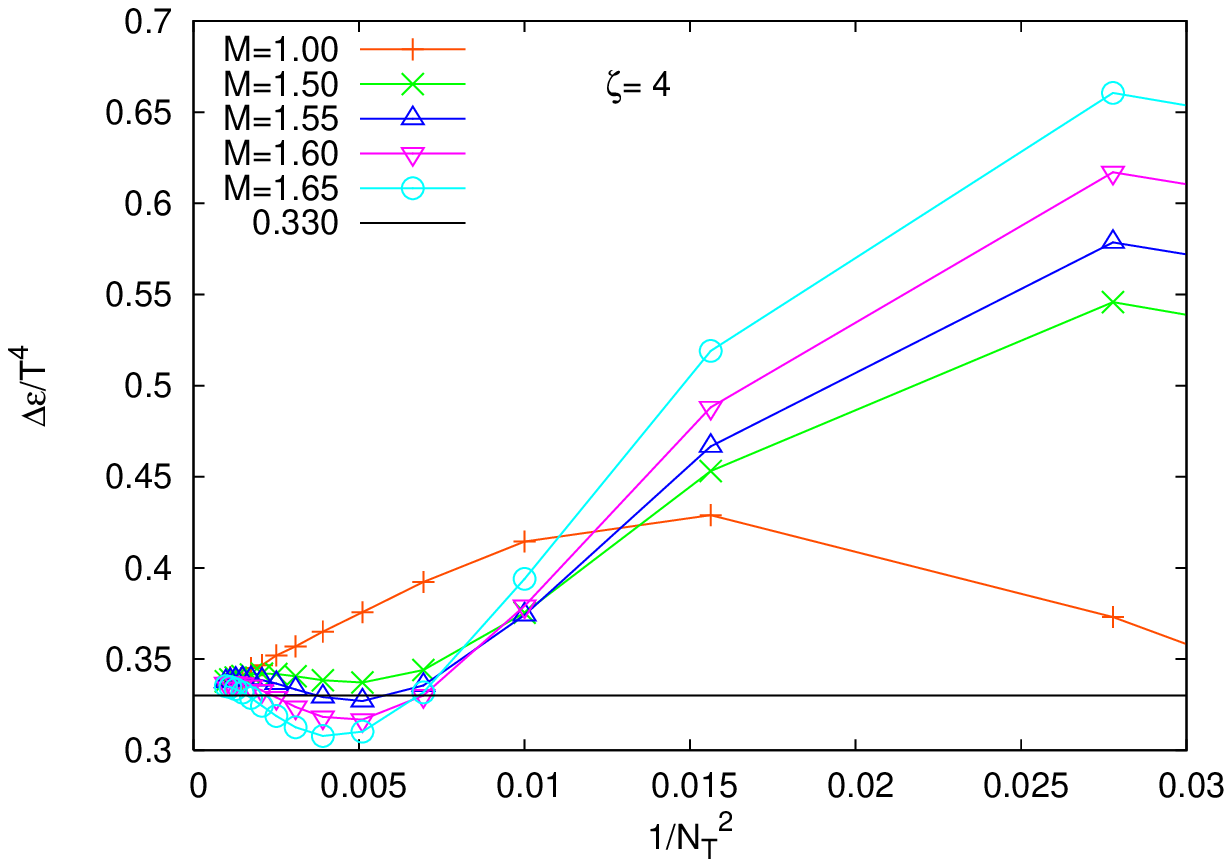}
\caption{The variation of the lattice $\Delta \epsilon/T^4$ with $1/N_T^2$
 for  $\hat{\mu}=0.5/N_T$(upper panel) and  $\hat{\mu}=0.8/N_T$(lower panel).}
\label{DelE}
\end{center}
\end{figure}

From the plots of the ratio of the $\Delta \epsilon/T^4$ on the lattice and in
the continuum (eq.(\ref{eqn:dem})), shown in Figure \ref{DelE} we again 
conclude that the continuum limit is reached for $N_T \ge 12$ for both the 
cases for essentially all $M$, with the $1.5<M<1.6$ region displaying the 
smallest deviations in the region $N_T > 12$ as in the $\hat{\mu=0}$ case 
in section II.A.
Moreover the results for $\Delta \epsilon$ again appear about 1.6-1.8 times
larger on the lattices with $N_T=4-8$ while the susceptibility is close to
twice the continuum result.
\begin{figure}
\begin{center}
 \includegraphics[scale=0.7]{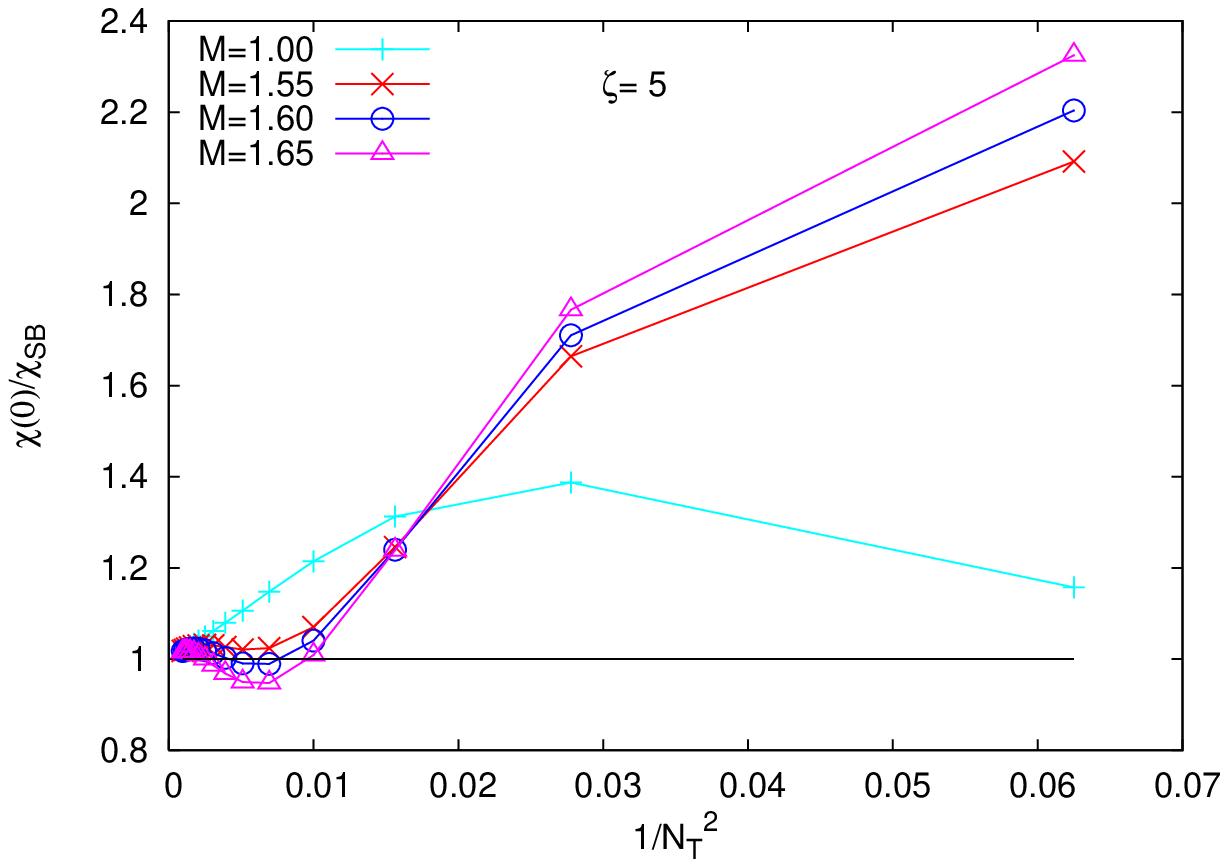}
\includegraphics[scale=0.7]{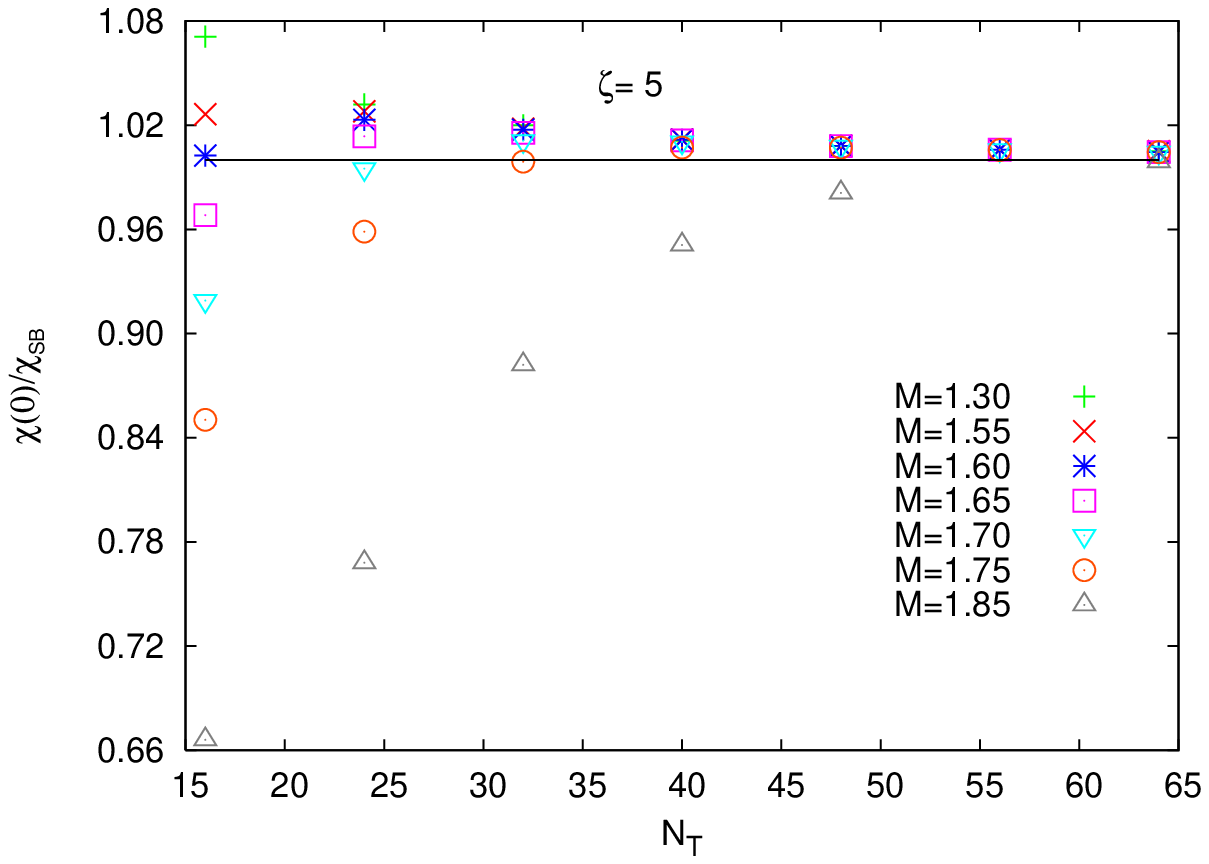}
\caption{The variation of $\chi(0)/\chi_{SB}$ vs $1/N_T^2$(upper panel)
and vs $N_T$(lower panel) for different M and $\zeta=5$.}
\label{susc}
\end{center}
\end{figure}

In the upper panel of Figure \ref{susc}, we display the ratio of
$\chi(\mu=0,T)$ on the lattice with the corresponding continuum value from
eq.(\ref{eqn:sus}). One sees again a similar pattern as for the energy density.
As in eq.(\ref{eqn:laten}), the susceptibility calculated on the lattice will
also have a form,
\begin{equation}
 \chi(0)=\frac{B^\chi}{N_T^2}+\frac{C^\chi(M)}{N_T^4}+\frac{D^\chi(M)}{N_T^6}+...~,~
\end{equation}
where the only difference is the absence above of a constant term like $A$.
Keeping only the first term, one will again get the effective $B^\chi$ to become
$M$-dependent; its deviation from 1/3 will be a measure of the finite lattice
spacing effects.  The filled circles in Figure \ref{sbdv} display these artifact
effects as a function of $M$ which were obtained by assuming a constant
behaviour in the range $18 \le N_T \le 32$.  The absence of a dominant term like
$A$ in the equation above allowed us to re-do the fit with the inclusion of the
next term for each $M$.  We found that the resultant $B^\chi$ is already
$M$-independent and close to 1/3 in each case.  Moreover the $C^\chi$ changed
with $M$ substantially and was smallest for $M=1.6$.  From all these fits, it
also emerged that by $N_T=64$ the contribution of the $C^\chi$-term becomes
negligible.  The lower panel of Figure \ref{susc} exhibits the results of our
attempt to verify this by extending the computations to larger lattices.  One
does, indeed, find a convergence to the continuum result irrespective of the
value of M from lattice sizes of $320^3\times64$.  Note that one finds very
similar effects of finite lattice spacing for both the susceptibility and the
energy density at $\mu=0$ in Figure \ref{sbdv}, with $M \sim 1.6$ emerging as a
good choice for calculations on lattices with small $N_T$ due to smallest 
values of the correction terms.

\section{Massive overlap fermions}

While we restricted ourselves to the thermodynamics of massless overlap
fermions, most of our treatment goes through for the massive fermions as well.
In this section we outline this for the $\mu=0$ case.  For the sake of novelty,
we use an alternative way of doing the computation.  The overlap operator for
fermions of mass $m$ is written as
\begin{equation}
 D_{ov}=(1+\frac{m}{2M})+(1-\frac{m}{2M})sgn(\gamma^5 D_W)~.~
\end{equation}
The eigenvalues of the overlap-Dirac operator change from $\lambda_{\pm}$ in
eq.(\ref{eqn:lam}) to $\lambda_{\pm}\rightarrow\lambda_{\pm}(1-m/2M)+m/M$. As
a result the energy density modifies from eq.(\ref{eqn:x3}) to
\begin{equation}
\label{eqn:x4}
\epsilon = \frac{2}{N^3 a^3 N_T} \sum_{p_j,p_4}\frac{\alpha(h^2 \frac{\partial
h_5}{\partial a_4}-h_5 h_4 \frac{\partial h_4}{\partial a_4})}{
(h^2+h_5^2)(\gamma\sqrt{h^2+h_5^2}-\alpha h_5)}~,
\end{equation}
where 
\begin{equation*}
 \alpha=2\left(1-\frac{m^2a^2}{4M^2}\right)~~~{\rm and}~~~
 \gamma=2\left(1+\frac{m^2a^2}{4M^2}\right)~.
\end{equation*}
Substituting the values of $h_4$, $h_5$ and their derivatives, one obtains
\begin{eqnarray}
\label{eqn:x5}
&&\epsilon a^4 = \frac{2\alpha}{N^3 N_T}\sum_{p_j,n}\nonumber\\
&[ &\frac{(1-\cos\omega_n) (f+ \sin^2\omega_n)}{(d+2g\cos\omega_n)(\gamma^2(d+2g\cos\omega_n)
-\alpha^2(g+\cos\omega_n)^2)} \nonumber\\ 
&+&\frac{\sin^2\omega_n(g+\cos\omega_n)}{(d+2g\cos\omega_n)(\gamma^2(d+2g\cos\omega_n)
-\alpha^2(g+\cos\omega_n)^2)}]\nonumber\\
 &\times& \left[\gamma\sqrt{d+2g \cos\omega_n} +\alpha(g+\cos\omega_n)\right]~.
\end{eqnarray}
Note that setting $m=0$, reduces $\alpha=\gamma$.  Substituting in the equation
above, and using the relation $d = g^2 +f +1$, it becomes identical to the
expression in eq.(\ref{eqn:xxx}), as expected.

One can again use the same contour method for evaluating the energy density. 
By comparing with eq.(\ref{eqn:xxx}),  the functions $F_1$ and $F_2$ can be 
identified as the two terms obtained by removing the second pair of brackets 
of eq. (\ref{eqn:x5}).  The poles (and branch cuts) of these functions can be
seen to be the same except that the poles defined by 
$\omega=\pm i~ \text{sinh}^{-1}\sqrt{f}$ are now given by 
\begin{equation}
\label{eqn:pole}
\cos(\omega) = y \pm z ~,~
\end{equation}
where $y$ and $z$ are defined as
\begin{eqnarray}
y &=& g (\frac{\gamma^2}{\alpha^2}-1)\\
z &=& \frac{\gamma}{\alpha}\sqrt{g^2 (\frac{\gamma^2}{\alpha^2}-1)+f+1}~.~
\end{eqnarray}
As proved in Appendix \ref{massiveOVL}, $z-y>1$, making abs$(\cos \omega) >1$ or $\omega$
purely imaginary. The pole $\omega= i \text{cosh}^{-1} (y+z) $ lies on the
imaginary axis while that for  $\omega= i \text{cosh}^{-1} (y-z) $ lie on
parallel lines shifted by $\pm \pi$.  The choice of contour can be made similar to that in Figure \ref{cont}, allowing only the former to contribute to the
energy density. 

Setting $m=0$, it is easy to verify that this approach also yields precisely
the result in eq.(\ref{eqn:lat}) by selecting the contour as in the upper half
plane of Figure \ref{cont}.  Its analog for $m \neq 0$ by can be obtained by
computing the residue at the pole defined by eq.(\ref{eqn:pole}).  Instead of
giving the full expression again, we only indicate  how the results in the
continuum limit arise.  The pole  positions can be computed to be 
\begin{equation}
 \cos\omega=\alpha_1=(1+\frac{a^2(\vec{p}^2+m^2)}{2})~,~
\end{equation}
and  denoting by $m'=m(M-2)/M$
\begin{equation}
\cos\omega=\alpha_2=-(1+\frac{a^2(\vec{p}^2+m'^2)}{2})~,~
\end{equation}
where $\vec{p}^2=p_1^2+p_2^2+p_3^2$.  The pole at $\alpha_1$
has, at order $a$, the residue
\begin{equation*} \text{Res}~F_1(\alpha_1)= \pm \frac{a\sqrt{\vec{p}^2+m^2}}{2}
\end{equation*} 
Note that all the other poles, including the poles at $\alpha_2$, and the
branch cuts do not contribute to the contour integrals, as seen in Figure
\ref{cont}.  Therefore the energy density in the continuum comes out to be the
same as in eq.(\ref{eqn:sb}) but with $E=\sqrt{m^2+p_{1}^2+p_{2}^2+p_{3}^2}$.

\section{Summary}
Investigating the thermodynamics of QCD on lattice with fermions which possess
both the chiral symmetry and the flavour symmetry relevant to our world 
has important consequences for both the experimental aspects of the heavy ion
collisions and the theoretical aspects of the $\mu-T$ phase diagram. Staggered
fermions used in the bulk of the work so far are not adequate to resolve some of
these issues. Overlap fermions, while computationally more expensive, may
prove better in such studies in near future. 

We have presented analytical and numerical results on the the thermodynamics of
free overlap fermions in 4-D both for zero and nonzero (baryonic) chemical
potential by varying the irrelevant parameter M.  From the energy density
computed on the lattice in these cases, we showed that the expected continuum
limit is reached.  Generalizing the proposed action \cite{wet} for nonzero
$\mu$, we demonstrated  that the $\mu^2$-divergence in the continuum limit is
avoided for a class of functions $K(\hat{\mu})$ and $L(\hat{\mu})$ with
$K(\hat{\mu}) \cdot L(\hat{\mu}) = 1$; the choice $ \exp(\pm \hat{\mu})$ for
$K$,$L$ also satisfies this condition is therefore shown to be free of any
$\mu^2$-divergence.  In all these case, however, the chiral invarinace of
the action is lost for nonzero $\mu$.

While the sign function in the free overlap Dirac operator remains a constant
in computations for $\mu=0$, we pointed out that it becomes undefined on
lattices with odd number of temporal sites for $ \mu \ge \mu_c$, where the
value of $\mu_c$ depends on $M$. Our numerical computations were restricted to
smaller $\mu$-values.  The numerical results were mildly dependent on the
aspect ratio of the spatial and temporal direction but changed significantly as
a function of the irrelevant parameter $M$ of the overlap Dirac operator.  For
the choice of $ 1.5 \le M \le 1.6$, both the energy density and the quark
number susceptibility computed for $\mu=0$ exhibited the smallest deviations
from the ideal gas limit, as seen in Figure \ref{sbdv}.  As seen from Figures
\ref{zeta}, \ref{DelE} and \ref{susc}, lattice results approximate the
continuum well for lattices with 12 or more temporal sites, with typically a
factor $\approx$ 1.8 larger results for smaller lattices with 6-8 temporal
sites.  It would be interesting to check whether the optimum M-range is 
still the same in the presence of gauge fields.

\section*{Acknowledgments}
S.S would like to acknowledge the Council of Scientific and Industrial
Research(CSIR) for financial support through the Shyama Prasad Mukherjee(SPM)
fellowship and thank Saumen Datta for useful discussions.

\appendix 
\section{Energy density for $\mu = 0$}
\label{zeroMU}
\begin{figure}
\begin{center}
\includegraphics[scale=0.3]{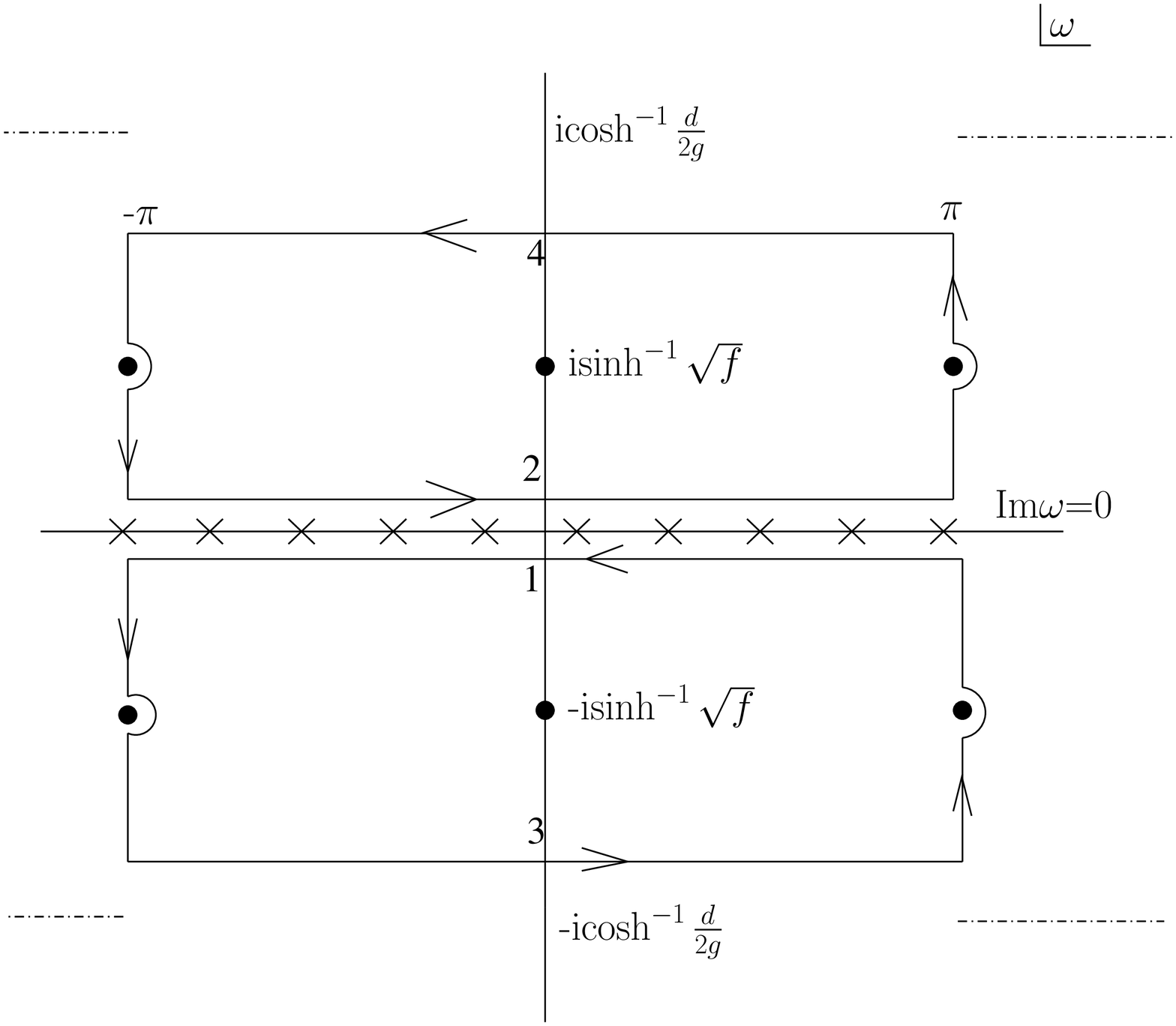}
\caption{The choice of contours for evaluating the $\omega$-sum in
eq.(\ref{eqn:xxx}). The dashed lines represent branch cuts. The crosses denote
the Matsubara frequencies $\omega_n$, while the filled circles denote the poles
of $F(\omega)$.}
\label{cont}
\end{center}
\end{figure}

Before we show the details of the energy density calculation, 
let us list certain useful relations amongst the quantities $g$,$d$,$b$ and 
$c$ introduced in eq.(\ref{eqn:aux}),~(which will be useful for the calculations below)~:
\begin{itemize} 
\item Since $ \cos(ap_j) \le 1$ for any $j$, $g<0$ for $M<1$,
\item $g^2 + f +1 = d \Longrightarrow d>0$ , since f is a sum of squares,
\item $d^2/4g^2 - f -1 =  (g^2-f-1)^2/2g^2  \\ \Longrightarrow d^2/2g^2> 1+f $,
\item $\text{cosh}^{-1}\frac{d}{2g}>\text{sinh}^{-1}\sqrt{f}$.  This follows
trivially from the line above. Since $ \frac{d^2}{4 g^2} > (1+f)$, it follows
$\frac{d}{2 g} > \sqrt{1+f}$ ( $\frac{d}{2 g} < -\sqrt{1+f}$, for $g < 0$).
Noting that $\text{cosh}(\text{sinh}^{-1}\sqrt{f}) = \sqrt{1+f}$, one has
$ \text{cosh}^ {-1}(\frac{d}{2g}) > \text{sinh}^{-1} \sqrt{f}$ 
($ \text{cosh}^ {-1}(\frac{d}{2g}) <  - \text{sinh}^{-1} \sqrt{f}$ for $g <0$).
\end{itemize}
The last line justifies the drawing of the contour in Figure \ref{cont} by 
avoiding the poles/cuts at $ \pm i~ \text{cosh}^ {-1}(\frac{d}{2g})$.

The summation over all Matsubara frequencies $\omega_n=\frac{(2n+1)\pi}{N_T}$ can be done using the
standard textbook\cite{rothe} method of contours in the complex $\omega$-plane.
For a general function $F(\omega)$, which in addition may, and does, depend on
other variables such as $p_j$, but this dependence will not be shown explicitly
below, the frequency sum therefore is,
\begin{equation}
\label{eqn:cont}
  \frac{2\pi}{N_T}\sum_{n}F(\omega_n)=\int_{\pi-i\epsilon}^{-\pi-i\epsilon}
\frac{F(\omega) d\omega}{e^{i\omega
N_T}+1}+\int_{-\pi+i\epsilon}^{\pi+i\epsilon}\frac{F(\omega) d\omega}{e^{i\omega
N_T}+1}~,~
\end{equation}
where the integrals are evaluated on the contour lines running parallel to the
real axis. The second integral can further be re-written as
\begin{equation}
\label{eqn:fsum}
\int_{-\pi+i\epsilon}^{\pi+i\epsilon}\frac{F(\omega) d\omega}{e^{i\omega N_T}
+1} =\int_{-\pi +i\epsilon}^{\pi +i\epsilon} F(\omega)
d\omega-\int_{-\pi+i\epsilon}^{\pi+i\epsilon}\frac{F(\omega)
d\omega}{e^{-i\omega N_T}+1}~.~ 
\end{equation}

The summand $F(\omega)$ in eq.(\ref{eqn:xxx}) can be split in to two terms,
\begin{equation*}
 F(\omega)=F_1(\omega)+F_2(\omega)~,~
\end{equation*}
with
\begin{eqnarray}
\nonumber
F_1(\omega)&=&\left(\frac{(1-\cos\omega) }{d+2g \cos\omega}+\frac{\sin^2\omega(g+\cos\omega)}{(d+2g \cos\omega)(f+\sin^2\omega)}\right )\\
&\times&(g+ \cos\omega)~,~
\end{eqnarray}
and
\begin{eqnarray}
\nonumber
F_2(\omega)&=&\left(\frac{(1-\cos\omega) }{d+2g \cos\omega}+\frac{\sin^2\omega(g+\cos\omega)}{(d+2g \cos\omega)(f+\sin^2\omega)}\right )\\
&\times&\sqrt{d+2 g \cos\omega}~.~
\end{eqnarray}
Both the functions  $F_i$ have a finite number of poles at $\omega=\pm i~
\text{sinh}^{-1}\sqrt{f}$ and $\pm m \pi\pm i~ \text{sinh}^{-1}\sqrt{f}$, where
m is an integer.
Furthermore, $F_1$ has poles for $\frac{d}{2g}>0$  at $\omega=\pm k \pi\pm
i~\text{cosh}^{-1}\frac{d}{2g}$ while $F_2$ has branch points at the same
locations.  Similarly, for $\frac{d}{2g}<0$ the poles (branch points) of $F_1$
($F_2$) are at $\pm i\text{cosh}^{-1}\frac{d}{2g}$.  In the rest of the complex
$\omega$ plane both the functions are analytic. In view of these properties the
contours in eq.(\ref{eqn:cont}) can be deformed to the contours shown in Figure
\ref{cont}. We chose each contour such that it lies below (above) the cut in
the upper (lower) half of the plane.  As shown above,
$\text{cosh}^{-1}\frac{d}{2g}>\text{sinh}^{-1}\sqrt{f}$. Defining therefore
$2\eta=\text{cosh}^{-1}\frac{d}{2g}-\text{sinh}^{-1}\sqrt{ f}$ with $\eta > 0$,
the lines $3$ and $4$ are drawn through the  points $\mp(
i~\text{sinh}^{-1}\sqrt{ f} +  i\eta)$ respectively to avoid the cuts shown in
Figure \ref{cont}.

Consequently, the frequency sum in eq.(\ref{eqn:cont}) becomes 

\begin{eqnarray}
\label{eqn:enden}
\nonumber
\frac{2\pi}{N_T}\sum_{n}F(\omega_n)&=&-2\pi i\sum_{ {\rm Im}~\omega>0}\frac{\text{Res}~F(\omega)}{e^{-i\omega N_T}+1}\\
\nonumber
&+&2\pi i\sum_{ {\rm Im}~\omega<0} \frac{\text{Res}~F(\omega)}{e^{i\omega N_T}+1}\\
\nonumber
 &-&\int_3\frac{F(\omega) d\omega}{e^{i\omega N_T}+1}+\int_4 \frac{F(\omega) d\omega}{e^{-i\omega N_T}+1}\\
&+&\int_{-\pi +i\epsilon}^{\pi +i\epsilon} F(\omega) d\omega~.~
\end{eqnarray}
\noindent
Note that the line integrals along the vertical lines through $\pi$
and $-\pi$ cancel each other due to the periodicity of the function
$F(\omega)$. Indeed, in general for any function $G(\omega)$ satisfying 
the property, $G(\pi+i\eta)=G(-\pi+i\eta)$, the sum of integrals of 
$G(\omega)$ along opposite vertical paths of equal length through $-\pi$ and 
$\pi$ is identically zero.

The residues of the function $F_{1,2}(\omega)$ at the poles $\omega=\pm
i~\text{sinh}^{-1}\sqrt{ f}$ are $\pm iR_1$ and $\pm iR_2$ respectively where
\begin{equation}
\label{eqn:res1}
R_1= R_2 = \frac{\sqrt{f}}{2\sqrt{1+f}} 
\end{equation}

Our choice of the contour ensures that the poles at $\pm\pi\pm i~
\text{sinh}^{-1}\sqrt{f}$ do not contribute to the energy density. By taking
the limit $N_T\rightarrow\infty$ on the lattice, on finds that the last term of
eq.(\ref{eqn:enden}) gives the quartically divergent vacuum contribution in 
the continuum limit.  Defining the physical energy density by subtracting it
off, we have

\begin{eqnarray}
\label{eqn:lat}
\nonumber
\epsilon a^4& =&\frac{4}{N^3} \sum_{p_j}\left [\frac{\sqrt{f}}{\sqrt{1+f}}\right]\frac{1}{e^{ N_T\sinh^{-1}\sqrt{f}}+1}\\
&+&\epsilon_3+\epsilon_4~,~
\end{eqnarray}
 where $\epsilon_3$, $\epsilon_4$ terms come from from the line integrals
3 and 4 respectively.

\section{Energy density at $T\neq 0$ and $\mu \neq 0$}
\label{nonzeroTMU}
\begin{figure}
\begin{center}
 \includegraphics[scale=0.35]{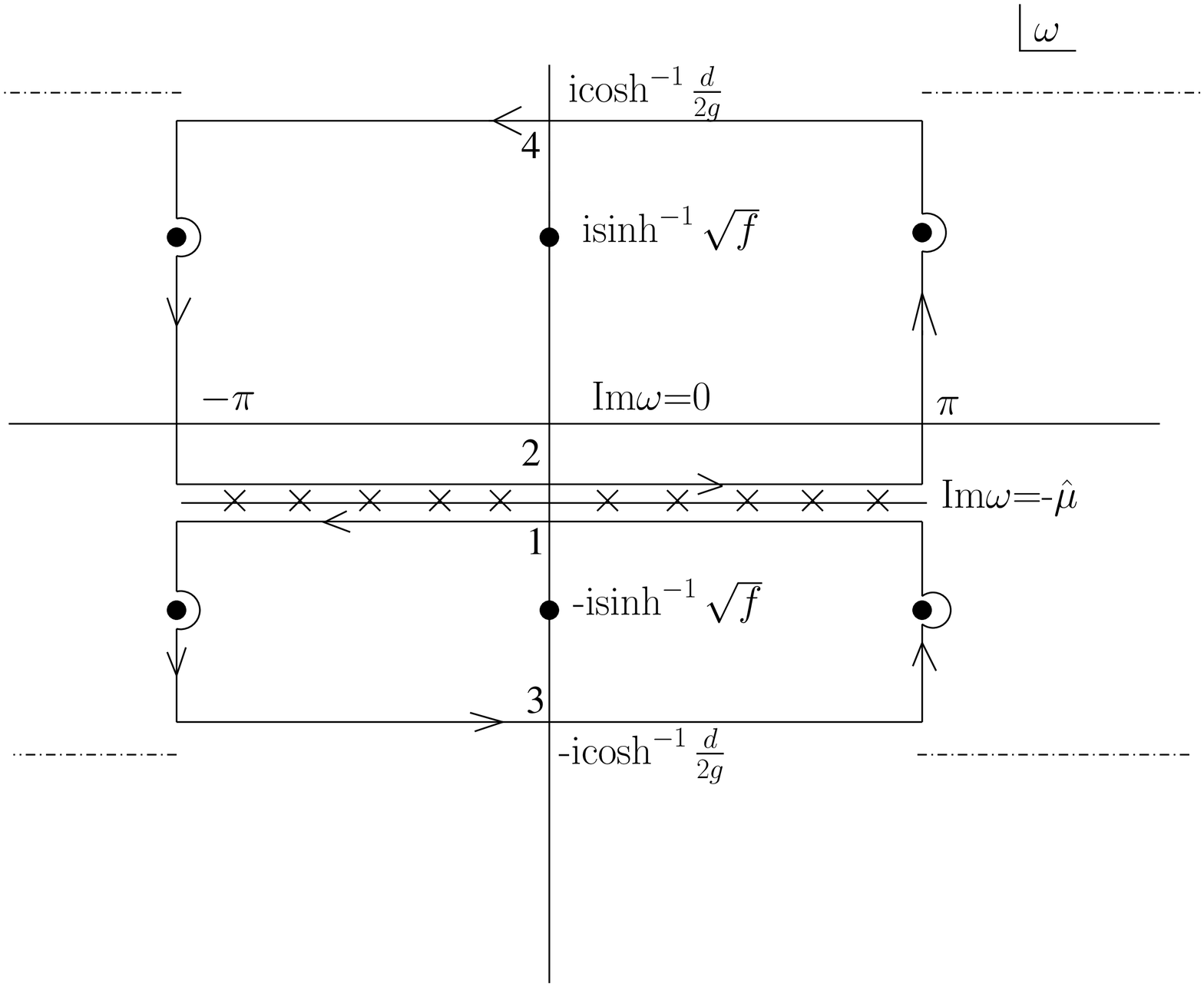}
\caption{The contours chosen for evaluation of the $\omega-$ sum for  
$ \hat{ \mu}< \sinh^{-1}\sqrt{f}$. Same notation as in
Figure \ref{cont}.}
\label{cont2}
\end{center}
\end{figure}

For evaluating the energy density, we revert back to the choice $K(\hat{\mu})=\exp(\hat{\mu})$
and $L(\hat{\mu})=\exp(-\hat{\mu} )$ which has $R=1$ and $ \theta = \hat{\mu}$.
 The physical part of the energy density on the lattice is calculated
by subtracting off the $\hat{\mu}=0 ~,~T=0$  contribution,  
 \begin{equation}
\label{eqn:finmu}
\epsilon a^4=\sum_{p_j}\frac{1}{N^3}\left[\frac{2}{ N_T}\sum_{n}F(\omega_n-i\hat{\mu})-\frac{1}{\pi}\int_{-\pi}^{\pi}F(\omega) d\omega\right]
\end{equation}
The further evaluation of the energy density  can be done, noting that compared
to the Figure \ref{cont}, the Matsubara frequencies $(2n+1)\pi/N_T$  in this
case are displaced along the imaginary axis by $i\hat{ \mu}$ in the lower half
plane with the choice of the function  $1/(\exp[i(\omega+i\hat{\mu}) N_T]+1)$.
The frequency sum can be replaced by line integrals as,
 \begin{eqnarray}
\label{eqn:sum1}
\nonumber
\frac{2\pi}{ N_T}\sum_{n}F(\omega_n-i\hat{\mu})&=&\int_{\pi-i\epsilon-i\hat{\mu}}^{-\pi-i\epsilon-i\hat{\mu}}\frac{F(\omega) d\omega}{e^{i(\omega+i\hat{\mu})N_T}+1}\\ 
&+&\int_{-\pi+i\epsilon-i\hat{\mu}}^{\pi+i\epsilon-i\hat{\mu}}\frac{F(\omega) d\omega}{e^{i(\omega+i\hat{\mu})N_T}+1}
 \end{eqnarray}
The choice of the contour will be analogous to that in Figure \ref{cont}, but
will depend on whether the pole $\omega=-i~\sinh^{-1}\sqrt f$ is above or below
the $\text{Im} \omega=-\hat{\mu}$ line. Therefore the frequency sum can be split into
two terms,
 \begin{equation}
\label{eqn:sum2}
\sum_{n}F(\omega_n-i\hat{\mu})=\sum_{n}\left[F_{<}(\omega_n-i\hat{\mu})
+F_{>}(\omega_n-i\hat{\mu})\right]
\end{equation} 
where $F_{>}~\text{and}~F_{<}$ are the functions with $\sinh^{-1}\sqrt f<
\hat{\mu}$ ($\sinh^{-1}\sqrt f > \hat{\mu}$) respectively. We have taken
$\hat{\mu}<\cosh^{-1}\frac{d}{2g}$ because we expect that in the continuum
limit the $\hat{\mu}$ to scale as the lattice spacing whereas the second term
to tend to infinity. For the case $\hat{\mu}<\text{sinh}^{-1}\sqrt{f}$, the 
contour was chosen as shown in Figure \ref{cont2}. 

\begin{figure}
\begin{center}
\includegraphics[scale=0.3]{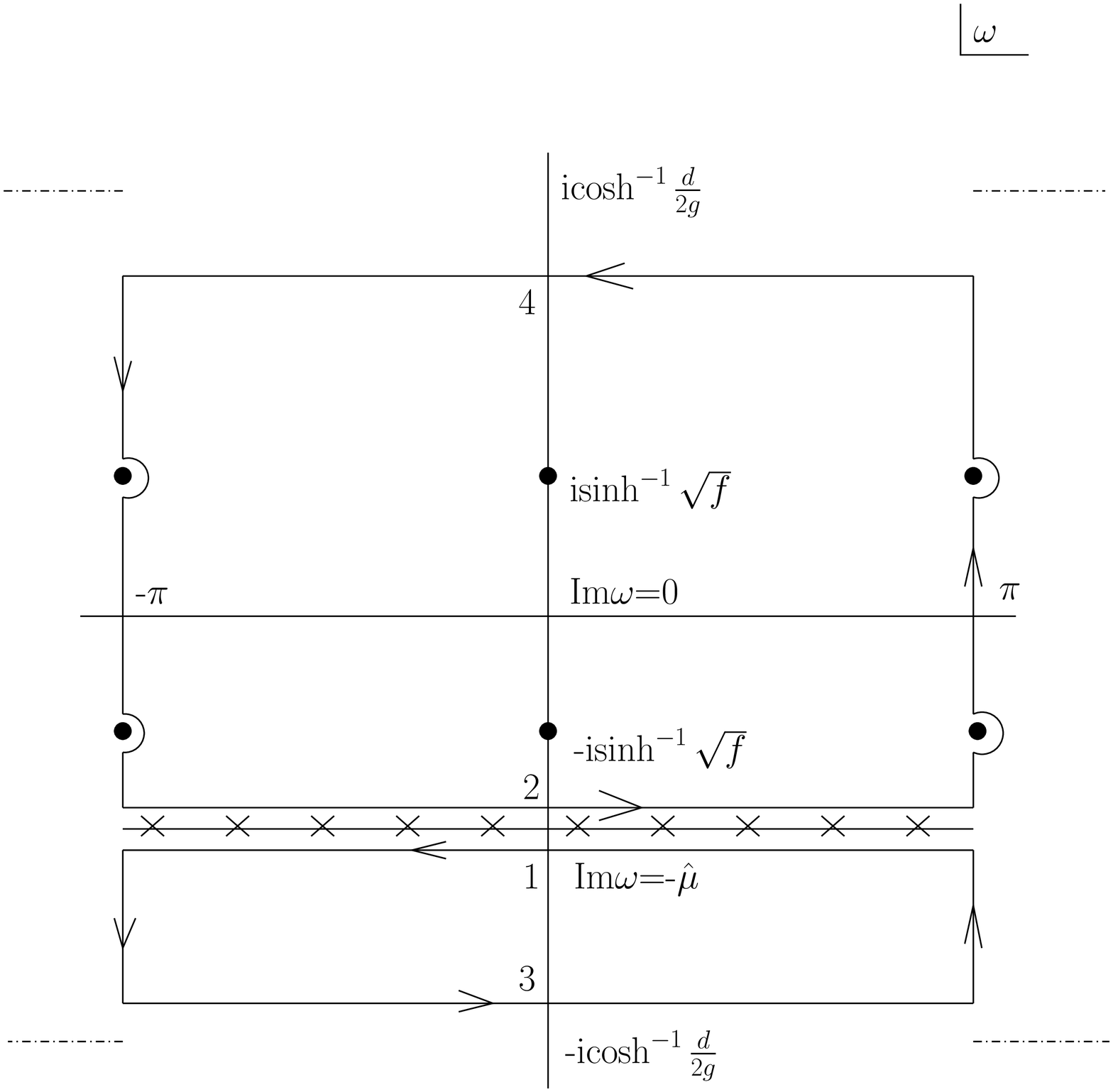}
\caption{The contours chosen for evaluation of the $\omega-$ sum for  $ \hat{
\mu}> \sinh^{-1}\sqrt{f}$. Same notation as in Figure \ref{cont}.}
\label{cont3}
\end{center}
\end{figure}

Using the standard trick of rewriting the second integral in
eq.(\ref{eqn:sum1}) as done in eq.(\ref{eqn:fsum})the frequency sum becomes
\begin{eqnarray}
\nonumber
 \frac{2\pi}{N_T}\sum_{n}F_{<}(\omega_n-i\hat{ \mu})&=&\Theta(\sinh^{-1}\sqrt f-\hat{\mu})\\
\nonumber
&\times&\left(-2\pi i\sum_{ Im \omega>0} \frac{\text{Res}~F(\omega)}{e^{-i(\omega+i\hat{ \mu}) N_T}+1}\right.\\
\nonumber
&+&2\pi i\sum_{ Im\omega<0} \frac{\text{Res}~F(\omega)}{e^{i(\omega+i\hat{ \mu}) N_T}+1}\\
\nonumber 
&-&\int_3\frac{F(\omega)}{e^{i(\omega+i\hat{ \mu}) N_T}+1} d\omega\\
\nonumber
&+&\int_4\frac{F(\omega)}{e^{-i(\omega+i\hat{ \mu}) N_T}+1} d\omega \\
&+&\left.\int_{-\pi}^{\pi}F(\omega)d\omega\right)
\end{eqnarray}
 The expression can be evaluated by
substituting the values of the residues of the function $F(\omega)$, which are
the sum of residues calculated in eq.(\ref{eqn:res1}). The integrals 3 and 4 
are along the lines $\text{Im}~\omega=\mp(\sinh^{-1}\sqrt f+\eta)$.
Representing them as $\epsilon_{3\mu,4\mu}$ respectively, we get
\begin{eqnarray}
\nonumber
 \frac{2\pi}{N_T}\sum_{n}F_{<}(\omega_n-i\hat{ \mu})&=&\Theta(\sinh^{-1}\sqrt f-\hat{\mu})\\
\nonumber
&\times& \left(\frac{4\pi~R_1}{e^{(\sinh^{-1}\sqrt{f}+\hat{ \mu}) N_T}+1}\right.\\
\nonumber
&+& \frac{4\pi~R_1}{e^{(\sinh^{-1}\sqrt{f}-\hat{ \mu}) N_T}+1}\\
\nonumber
&+&\left.\int_{-\pi}^{\pi}F(\omega)d\omega+\epsilon_{3\mu}+\epsilon_{4\mu}\right)\\
\end{eqnarray}
Similarly for $ \hat{ \mu}> \sinh^{-1}\sqrt{f}$ the frequency sum is replaced
by integrals along the contour as shown in Figure \ref{cont3}. 
There are no poles below the $\text{Im}~\omega=-\hat{\mu}$ line so the first integral
in eq.(\ref{eqn:sum1}) can be replaced by a line integral along the line 3. 
Following the same steps as discussed for the above case, the frequency sum reduces to

\begin{eqnarray}
\nonumber
 \frac{2\pi}{N_T}\sum_{n}F_{>}(\omega_n-i\hat{ \mu})&=& \Theta(\hat{ \mu}-\sinh^{-1}\sqrt f)\\
\nonumber
&\times&\left( \frac{4\pi~R_1}{e^{(\sinh^{-1}\sqrt f-\hat{ \mu}) N_T}+1} \right.\\
\nonumber
&+&\frac{4\pi~R_1}{e^{(\sinh^{-1}\sqrt f+\hat{ \mu}) N_T}+1} \\
&+&\left.\int_{-\pi}^{\pi}F(\omega) d\omega+\epsilon_{3\mu}+\epsilon_{4\mu}\right)~.~
\end{eqnarray}
Finally, the energy density on the lattice is obtained from 
eq.(\ref{eqn:finmu}) by substituting in the eq.(\ref{eqn:sum2}) the frequency
sums calculated above:
\begin{eqnarray}
\label{eqn:elatmu}
\nonumber
 \epsilon a^4&=&\frac{2}{N^3}\sum_{p_j}\left[\frac{\sqrt f}{\sqrt {1+f}}\frac{1}{e^{(\sinh^{-1}\sqrt f-\hat{ \mu}) N_T}+1}\right.\\
\nonumber
&+&\frac{\sqrt f}{\sqrt {1+f}}\frac{1}{e^{(\sinh^{-1}\sqrt f+\hat{ \mu}) N_T}+1}\\
&+&\left. \epsilon_{3\mu}+\epsilon_{4\mu} \right]
\end{eqnarray}

\section{}
\label{massiveOVL}
Here we prove the claim made in eq.(\ref{eqn:pole}).
\begin{itemize}
\item $y>0$ since $\gamma>\alpha$.
\item Let $\xi=\frac{\gamma^2}{\alpha^2}-1$, where $\xi>0$.
\item A little algebra shows that $z^2-(y+1)^2 = \xi[(g-1)^2+f]+f > 0$ which in turn implies the relation $z-y>1$.
\end{itemize}

\end{document}